\documentclass[10 pt, aps,reprint]{revtex4-1}

\draft 

\usepackage{amssymb}
\usepackage{amsmath,bm}
\pdfoutput=1
\usepackage{inputenc}
\usepackage[T1]{fontenc}
\usepackage{amsbsy}
\usepackage{physics}
\usepackage{dcolumn}
\usepackage{bm}
\usepackage{graphicx}
\usepackage{caption}
\usepackage{subfigure}
\usepackage{latexsym}
\usepackage{dsfont}
\usepackage{mathrsfs}
\usepackage{appendix}
\usepackage{bm}
\usepackage{subfigure}
\usepackage{hyperref}

\begin{document}


\title{Transparency in a periodic chain of quantum emitters strongly coupled to a waveguide}
\author{Debsuvra Mukhopadhyay$^{1}$ and Girish S. Agarwal$^{1,2}$}
\email{debsosu16@tamu.edu}
\email{girish.agarwal@tamu.edu}

\address{$^1$Institute for Quantum Science and Engineering, Department of Physics and Astronomy, Texas A$\&$M University, College Station, TX 77843, USA
\\$^2$Department of Biological and Agricultural Engineering, Texas A$\&$M University, College Station, TX 77843, USA}


\begin{abstract}
We demonstrate the emergence of transparent behavior in a chain of periodically spaced non-identical quantums emitters coupled to a waveguide, in the special case when the inter-atomic separation is a half-integral multiple of the resonant wavelength, i.e $kL$ is an integral multiple of $\pi$, with $k$ being the spatial frequency and $L$ the spatial periodicity. When equal but opposite frequency detunings are assigned in pairs to a system of even number of atoms, perfect transmission ensues. When the chain size is odd, a similar assignment leads to the disappearance of collective effects as the odd atom determines the spectral behavior. We also manifest the robustness of these features against dissipative effects and show, how the spectral behavior hinges significantly on the relative detunings between the atoms as compared to the decay rate. A key distinction from the phenomenon of Electromagnetically Induced Transparency (EIT) is that in the waveguide case, the presence of an intrinsic waveguide mediated phase coupling between the atoms strongly affects the transport properties. Furthermore, while reciprocity in single-photon transport does not generally hold due to the phase coupling, we observe an interesting exception for $kL=n\pi$ at which the waveguide demonstrates reciprocal behavior with regard to both the transmission and reflection coefficients.  \\


\end{abstract}
\pacs{Valid PACS appear here}

\maketitle 

\section{Introduction}
There is an ever-growing interest in single and few photon scattering from a one-dimensional (1D) continuum because of their possible applications in quantum information processing. This subject has been studied both theoretically \cite{c1,c2,c3,c4,c5,c6,c7,c8,c9,c10,c11,c12,c13,c14,c15,c16,c17,c18,c19,c20,c21,c22,c23,c24,c25,c26,c27,c28,c29,c30,c31,c32,c33,c34,c35,c36,c37} and experimentally \cite{c38,c39,c40,c41,c42,c43,c44,c45,c46,c47,c48} from a variety of perspectives and reviewed quite elaborately in Reference \cite{c49}. Some commonly used 1D waveguides are conducting nanowires \cite{c38,c39}, photonic crystal waveguides \cite{c47}, and superconducting microwave transmission lines \cite{c43,c44}. Collective effects emerging from a periodic array of two-level atoms coupled to a waveguide can lead to Fano minima in the reflection spectrum \cite{c28,c29,c37}, superradiant decays \cite{c47}, rearrangement of the optical band structure \cite{c50} and realization of Bragg mirrors \cite{c51,c52}. The role of spatial separation between the atoms in the context of photon scattering from a 1D continuum has been manifested in recent works \cite{c5,c31,c32,c51,c52,c53,c54,c55,c56,c57}. Very recently, multiple Fano interference channels due to the waveguide-mediated phase coupling between the atoms leading to the appearance of transparency points were demonstrated in \cite{c37}.

Theoretical analysis of photon scattering from a collection of disparately detuned atoms coupled to a waveguide is a hard problem in real space, because of the waveguide mediated phase coupling between the atoms. Here, we address the collective behavior of such a system by constraining the spatial periodicity to be an integral or half-integral multiple of the resonance wavelength. Working at this phase substantially simplifies the problem and provides insight into the various kinds of collective properties of the output radiation. Since our interest is primarily in the realizability of induced transparency, we demonstrate the emergence of new Fano minima in the reflection spectrum for a system of even number of emitters via appropriate choice of detunings. For an odd chain size, the system manifests an effective re-emergence of single-atom behavior. These are new effects that were not observed in \cite{c37}, as the atoms were all supposed to have identical transition frequencies. Our results are strong enough to withstand dissipative effects as long as the dissipative channel is weak compared to the waveguide channel and the atomic detunings are large compared to the decay rate. In other words, much like the phenomenon of EIT \cite{harris1991,eitreview}, while perfect transparency emerges in an ideal dissipation-free scenario, a fairly high degree of transparency can still be observed in the presence of weak dissipation. However, in the usual scenario for EIT, the system comprises of non-interacting atoms, whereas in a waveguide, an effective phase coupling between the emitters persists even at long spatial separations (of the order of a wavelength). Another important feature that surfaces at this choice of phase is the reciprocity in optical transport, i.e. with regard to both the reflected and transmitted amplitudes. More generally, the transport is insensitive to the order in which the atoms are placed in the chain. This is starkly different from the usual scenario, where the phase coupling makes the system strongly non-reciprocal. Reciprocity relations for a large class of one-dimensional systems were derived in \cite{rec1}. A general discussion on the subject of Lorentz reciprocity for lossless systems can be found in \cite{recrev}.

The paper is organized as follows: in Sec. \ref{s2}, we review the transport model for a single photon through a waveguide coupled to a chain of non-identical emitters. In Sec. \ref{s3}, we demonstrate how one can observe transparent behavior for an atomic chain consisting of two disparately detuned atoms, when $kL$ is chosen to be an integral multiple of $\pi$. In Sec. \ref{s4}, we write down the analytical forms for the reflection and transmission coefficients for an arbitrary chain size and generalize some of the observations to the case of an even chain size. Concurrently, we indicate, for an odd chain size, how one can effect a complete suppression of collective effects so as to recover single-atom behavior. In Sec. \ref{s5}, we argue that the commutativity between the transfer matrices leads to reciprocal transport properties Finally, in Sec. \ref{s6}, we probe the effect of including radiative decay into nonwaveguide modes on the spectral characteristics, with particular emphasis on the observability of transparent behavior. Section \ref{s7} summarizes the key ideas explored.

\section{Single-photon transport in a waveguide coupled to $N$ non-identical emitters} \label{s2}
 
\begin{figure}[!t]
 \captionsetup{justification=raggedright,singlelinecheck=false}
\centering
\includegraphics[scale=0.80]{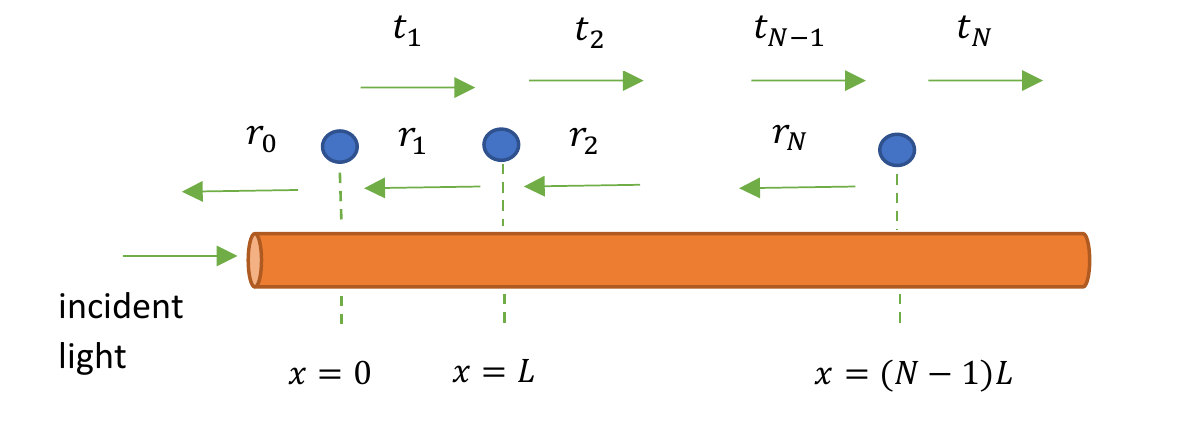} 
\caption{\small  {Atomic array coupled to a waveguide.}} \label{f1} 
\end{figure}

For a periodic 1-D array of $N$ two-level emitters evanescently coupled to a 1-D continuum (Fig. \ref{f1}), when the atomic transition frequency far exceeds the waveguide cutoff frequency, one can write down the Hamiltonian of the system in real space as:
\begin{align}
\mathcal{H}=i\hbar v_g\int_{-\infty}^{\infty}\dd{x}\bigg(a_L^{\dagger}(x)\dfrac{\partial a_L(x)}{\partial x}-a_R^{\dagger}(x)\dfrac{\partial a_R(x)}{\partial x}\bigg) \notag \\+\hbar\sum_{j=1}^{N}(\omega_j-i\Gamma_0)\ket{e}_j\bra{e}+ \notag \\ \hbar\sum_{j=1}^N\bigg[\{\mathcal{J}a_L(x_j)+\mathcal{J}a_R(x_j)+\text{h.c.}\}\ket{e}_j\bra{g}\bigg],
\end{align}
where $a_L(x)$ and $a_R(x)$ describe the real-space Bosonic operators corresponding to the left and the right propagating fields, $\omega_j$  corresponds to the transition frequency of the $j^{\text{th}}$ atom and $x_j=(j-1)L$ to its location along the waveguide, $v_g$ is the group velocity of the propagating photon and $\mathcal{J}$ denotes the coupling strength between the propagating field and any of the emitters. $\Gamma_0$ denotes the rate of spontaneous emission into all modes outside of the waveguide continuum, assumed equal for all the atoms. We disregard the dipole-dipole interaction (DDI) between the atoms by assuming the interatomic separation to be comparable or larger than the resonance wavelength.

The scattering eigenstate has the form
\begin{align}
\ket{\mathcal{E}_k}=\int_{-\infty}^{\infty}\dd{x}\bigg[\phi_{kL}(x)a_L^{\dagger}(x)+\phi_{kR}(x)a_R^{\dagger}(x)\bigg]\ket{\Psi} \notag\\
+\sum_{j=1}^Nc_k^{(j)}\ket{0;e_j},
\end{align}
where $\ket{\Psi}$ refers to the state with the field in vacuum and all atoms in the ground state, and $\ket{0;e_j}$ to the one where the field is still in vacuum but only the $j^{\text{th}}$ atom has been excited. $\phi_{kL}$ and $\phi_{kR}$ denote the wavefunctions corresponding to left- and right-propagating photonic excitations respectively. The Schr$\ddot{\text{o}}$dinger equation $\mathcal{H}\ket{\mathcal{E}_k}=\hbar v_g k\ket{\mathcal{E}_k} $ leads to a system of ODEs for the various probability amplitudes:
\begin{align}
\bigg(-iv_g\dfrac{\dd}{\dd x}-v_gk\bigg)\phi_{kR}(x)+\mathcal{J}\sum_{j=1}^{N}c_k^{(j)}\delta(x-x_j)&=0, \notag\\
\bigg(iv_g\dfrac{\dd}{\dd x}-v_gk\bigg)\phi_{kL}(x)+\mathcal{J}\sum_{j=1}^{N}c_k^{(j)}\delta(x-x_j)&=0, \notag\\
(\Omega_j-v_gk)c_k^{(j)}+\mathcal{J}\phi_{kL}(x_j)+\mathcal{J}\phi_{kR}(x_j)&=0,
\end{align}

where $\Omega_j=\omega_j-i\Gamma_0$. For a wave incident from the left, one can solve these ODEs subject to the boundary condition that $\phi_{kL/kR}(x_j)=\frac{1}{2}[\phi_{kL/kR}(x_j^+)+\phi_{kL/kR}(x_j^-)]$ and also the discontinuity imposed on the wavefunctions due to the the delta-function source, i.e. $-iv_g[\phi_{kR}(x_j^+)-\phi_{kR}(x_j^-)]+\mathcal{J}c_k^{(j)}=iv_g[\phi_{kL}(x_j^+)-\phi_{kL}(x_j^-)]+\mathcal{J}c_k^{(j)}=0$. The solutions take the form

\begin{align}
\phi_{kL}(x)&=\begin{cases} r_1e^{-ikx}, & x<0 \\
r_{j+1}e^{-ik(x-jL)}, & (j-1)L
<x<jL\\
0, & x>(N-1)L
\end{cases}\hspace{1mm},
\end{align}

\begin{align}
\phi_{kR}(x)&=\begin{cases} e^{ikx}, & x<0 \\
t_{j}e^{ik(x-jL)}, & (j-1)L<x<jL\\
t_Ne^{ik(x-NL)}, & x>(N-1)L
\end{cases}\hspace{1mm},
\end{align}

which subsequently lead to a system of simultaneous equations involving the transmission and reflection coefficients and the atomic excitation amplitudes. Eliminating the excitation amplitudes from the system engenders in a recursive matrix relation

\begin{align}
\begin{bmatrix}
r_j\\
t_{j-1}\\
\end{bmatrix}=\mathcal{L}_j\begin{bmatrix}
r_{j+1}\\
t_j\\ \end{bmatrix},
\end{align}
where
\begin{align}
\mathcal{L}_j=\begin{bmatrix}
e^{ikL}(1-i{\delta^{-1}_{k (j)}}) & -ie^{-ikL}\delta^{-1}_{k (j)}\\
ie^{ikL}\delta^{-1}_{k (j)} & e^{-ikL}(1+i\delta^{-1}_{k (j)}) \\
\end{bmatrix}, \label{E1}
\end{align}

with ${\delta_{k (j)}}=\dfrac{\omega_{k}-\Omega_j}{\Gamma}$ and $\Gamma=\dfrac{\mathcal{J}^2}{v_g}$ . Iterative use of this relation, conjoined with appropriate boundary conditions, yields the reflection and transmission coefficients:
\begin{align}
r=\dfrac{\bigg(\prod_{j=1}^N\mathcal{L}_j\bigg)_{12}}{\bigg(\prod_{j=1}^N\mathcal{L}_j\bigg)_{22}}, \hspace{5mm}t=\dfrac{e^{-ikNL}}{\bigg(\prod_{j=1}^N\mathcal{L}_j\bigg)_{22}}. \label{E2}
\end{align}

Because of the presence of the phase factors $e^{\pm ikL}$ and the differential detunings assigned to the emitters, it is analytically hard to find out compact expressions for the above. The matrix product, however, becomes simple to evaluate for $kL=n\pi$, $n$ being a natural number. 

\section{Transparency and collective behavior for a two-atom system} \label{s3}

Let us first consider the simpler scenario of two differentially detuned atoms in a waveguide and define the mean laser detuning $\overline{\Delta}=\omega_k-\frac{1}{2}(\omega_1+\omega_2)$ and relative atomic detuning $s=\omega_1-\omega_2$.  It turns out that while $[\mathcal{L}_1\mathcal{L}_2]_{22}$ is symmetric in $s$, $[\mathcal{L}_1\mathcal{L}_2]_{12}$ is not, owing to the phase coupling between the emitters mediated by the waveguide. In other words, in view of Eq. \ref{E2}, even though transmission is perfectly reciprocal, reflection is not. For this system, the reflection and transmission coefficients reduce to
\begin{align}
r&=\dfrac{-i\Gamma[(e^{i\alpha}+1)(\overline{\Delta}+i\Gamma_0)-(e^{i\alpha}-1){s}/{2}]-\Gamma^2(e^{i\alpha}-1)}{[\overline{\Delta}+i(\Gamma+\Gamma_0)]^2+\Gamma^2e^{i\alpha}-(s/2)^2}, \notag\\
t&=\dfrac{(\overline{\Delta}+i\Gamma_0+s/2)(\overline{\Delta}+i\Gamma_0-s/2)}{[\overline{\Delta}+i(\Gamma+\Gamma_0)]^2+\Gamma^2e^{i\alpha}-(s/2)^2}, \label{E4}
\end{align}

where we have taken $\alpha=2kL$. We note, however, in the special case $kL=n\pi$, that the above expressions turn symmetric in $s$, thereby leading to reciprocity in the transport properties. In subsequent considerations, we analyse the results pertaining to this special choice of phase.

\begin{figure}[!t]
 \captionsetup{justification=raggedright,singlelinecheck=false}
\centering
\includegraphics[scale=0.65]{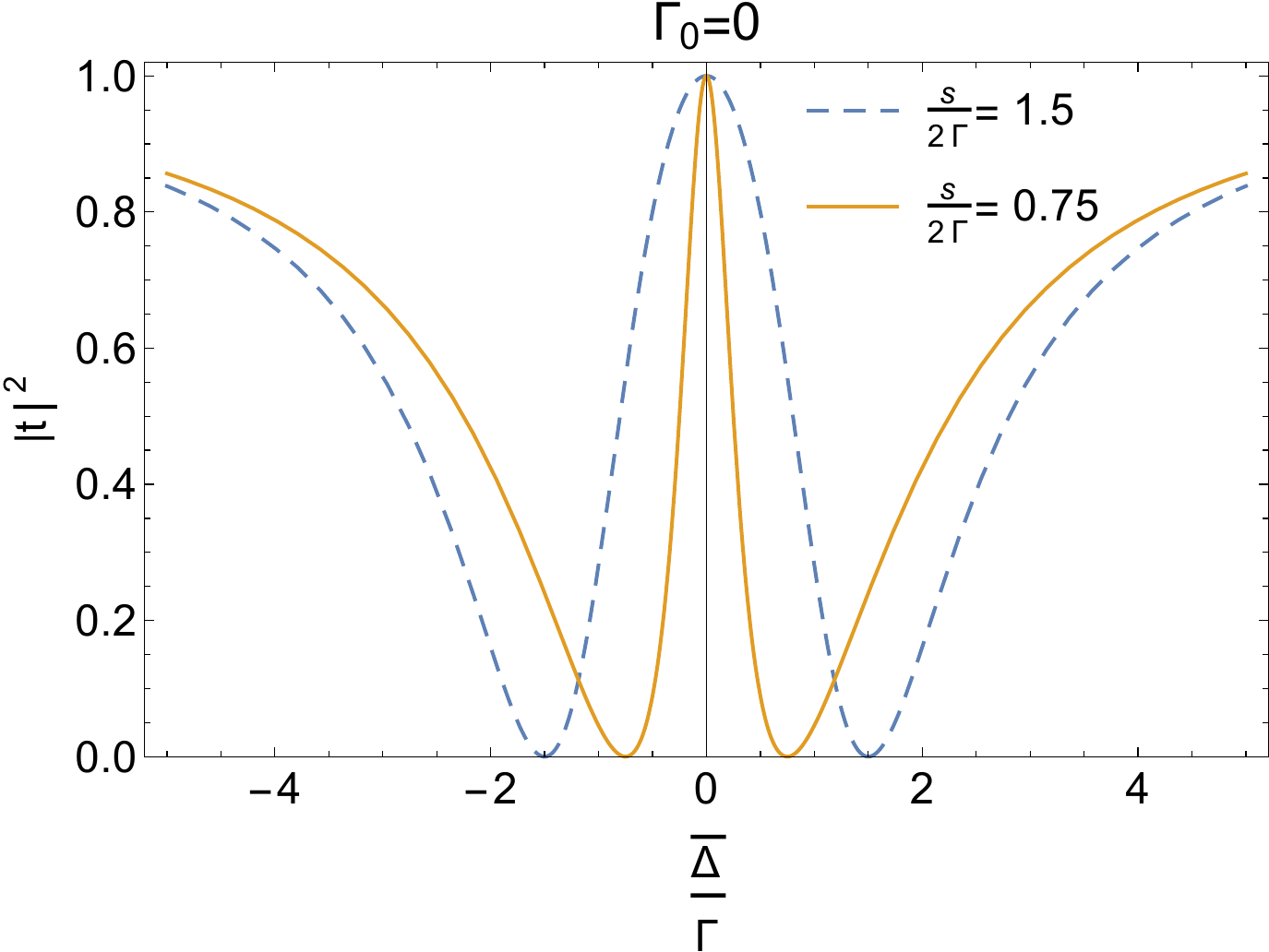} 
\caption{\small  {Transmission for a two-atom system without decay for a couple of values of $s=\omega_1-\omega_2$ and with $kL=n\pi$. Perfect transparency is observed at $\omega_k=\frac{1}{2}(\omega_1+\omega_2)$ (zero mean detuning), unless $\omega_1=\omega_2$, in which case, the system is perfectly reflecting at zero detuning. The two roots of the transmission come closer as the atomic frequencies approach each other.}} \label{figure2} 
\end{figure}

\begin{figure}[!t]
 \captionsetup{justification=raggedright,singlelinecheck=false}
\centering
\includegraphics[scale=0.65]{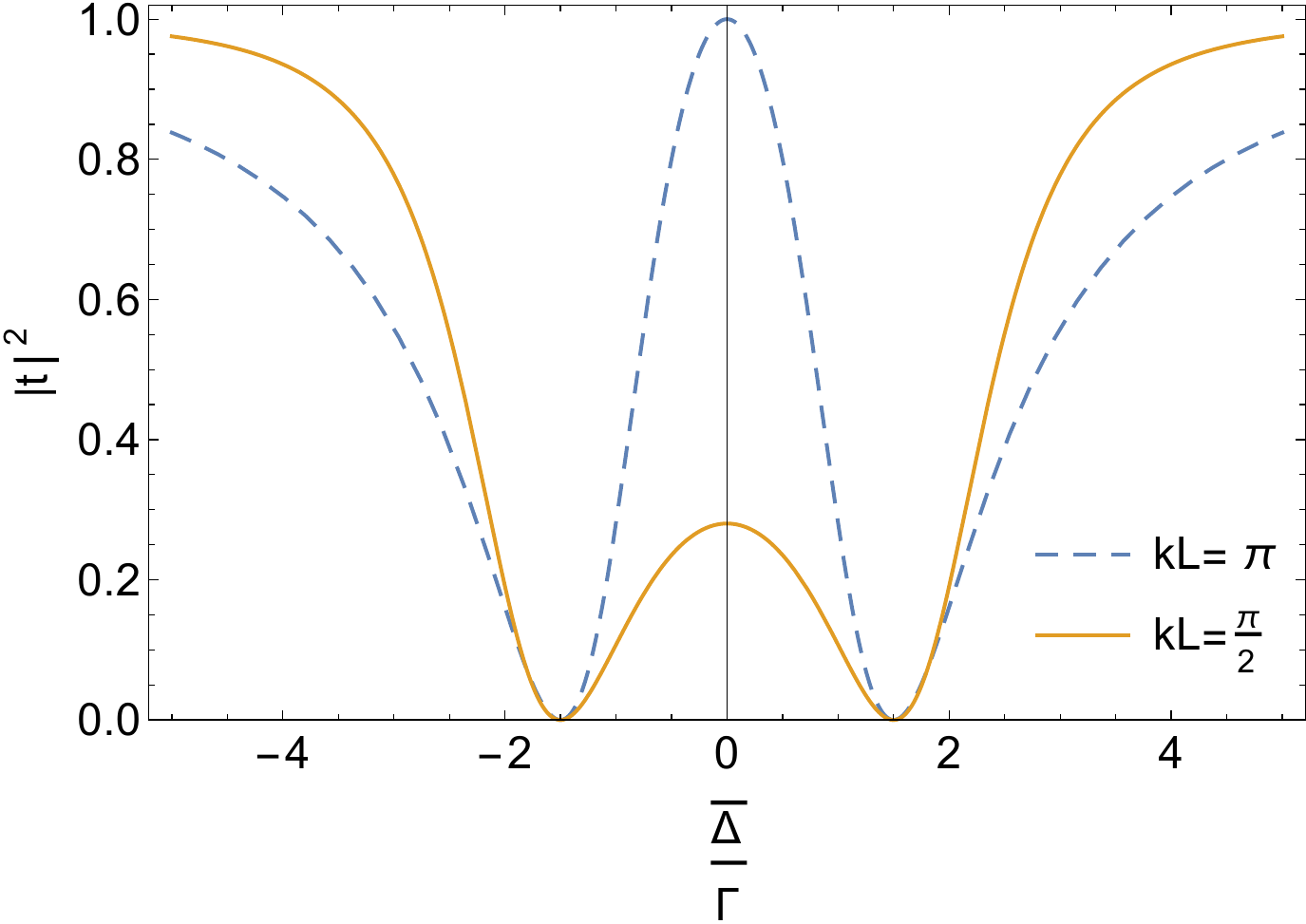} 
\caption{\small  {Comparison of transmission spectra corresponding to $kL=\pi$ and $kL=\frac{\pi}{2}$, with $s=1.5\Gamma$. The transmission peak attains unit height for $kL=\pi$ whereas it is much shorter than unity for $kL=\frac{\pi}{2}$.}} \label{f3} 
\end{figure}

We also assume an idealized scenario where $\Gamma_0$ can be ignored. The effect of $\Gamma_0$ on the transmission is studied later in Sec. \ref{s6}. With $\Gamma_0$ set to $0$, one observes an emergence of transparent behavior when the two atoms are equally detuned, albeit in opposite directions, with respect to the laser frequency. In other words, $t$ becomes unity when $\overline{\Delta}$ equals zero, or $\omega_k-\omega_1=\omega_2-\omega_k$, i.e. for a pair of antisymmetrically detuned emitters. It follows, from Eq. \ref{E4} and the plots in Fig. \ref{figure2}, that there is a transmission peak at $\overline{\Delta}=0$, while there are two roots of the profile at $\overline{\Delta}=\frac{s}{2}$ (or $\omega_k=\omega_1$) and $\overline{\Delta}=-\frac{s}{2}$ (or $\omega_k=\omega_2$) corresponding to perfect reflection.  The peak has unit height in the absence of decay, signifiying transparent behavior. The height of this peak is strictly less than unity for any other choice of phase, as can be verified from \ref{E4} (for instance, in the specific scenario, when $kL=\frac{n\pi}{2}$ with odd $n$, the height of this peak is $\bigg[\dfrac{s^2}{s^2+8\Gamma^2}\bigg]^2$ - see Fig. \ref{f3}). For a sufficiently small yet non-zero value of $\abs{\omega_1-\omega_2}$, one finds a very narrow window of size $s$ over which the system is capable of demonstrating both opacity as well as transparency. As $s\rightarrow 0$, the two roots come progressively closer. Fig. \ref{figure2} illustrates this scenario for various choices of $\abs{\omega_1-\omega_2}$. 

We also note, in passing, that the poles of the transmission and reflection demonstrate features remindful of level attraction. These poles occur at 
\begin{align}
\overline{\Delta}_{\pm}^{(\text{p})}=-i\Gamma\pm\sqrt{\bigg(\frac{s}{2}\bigg)^2-\Gamma^2}\hspace{0.5mm}.
\end{align}
Level attraction is typically observed between the normal modes of two coupled oscillators when one of the bare modes has negative energy and the modes have comparable decay rates. When the coupling strength equals or exceeds a critical value, the level separation vanishes. As a direct analogy, we see, in our case, that the real parts of the transmission poles coincide and become $0$ in the region $\frac{s}{2}\leq\Gamma$, while the imaginary parts expand and shrink respectively. The point of transition where the coupling equals this critical value is referred to as an exceptional point where the complex eigenfrequencies coincide \cite{E1}. Realizing level attraction has been quite a challenge from an experimental perspective and consequently, there is burgeoning interest in level attraction and phenomena occuring in the vicinity of exceptional points. Recently, level attraction has been observed in a variety of systems and topological behavior around an exceptional point has also been explored \cite{L1,L2,L3,L4,L5,L6,L7}.

In the waveguide case with two atoms, we see that the transmission has zeros at
\begin{align}
\overline{\Delta}_{\pm}^{(\text{r})}=\pm\frac{s}{2},
\end{align} 
which simultaneously determine the peaks of the reflection spectrum. These zeros are close to the the real parts of the poles when $\frac{s}{2}\gg\Gamma$. In the complementary regime, when $\frac{s}{2}$ is comparable to $\Gamma$, the real parts of the poles become small compared to the respective imaginary parts, as a consequence of which, the resolution between the two levels (or the two poles) becomes difficult. This problem of resolution arises fundamentally because $\Gamma$ not only appears in the discriminant of the poles, but also acts as a natural broadening term.\\

\begin{figure}[!t]
 \captionsetup{justification=raggedright,singlelinecheck=false}
\centering
\includegraphics[scale=0.65]{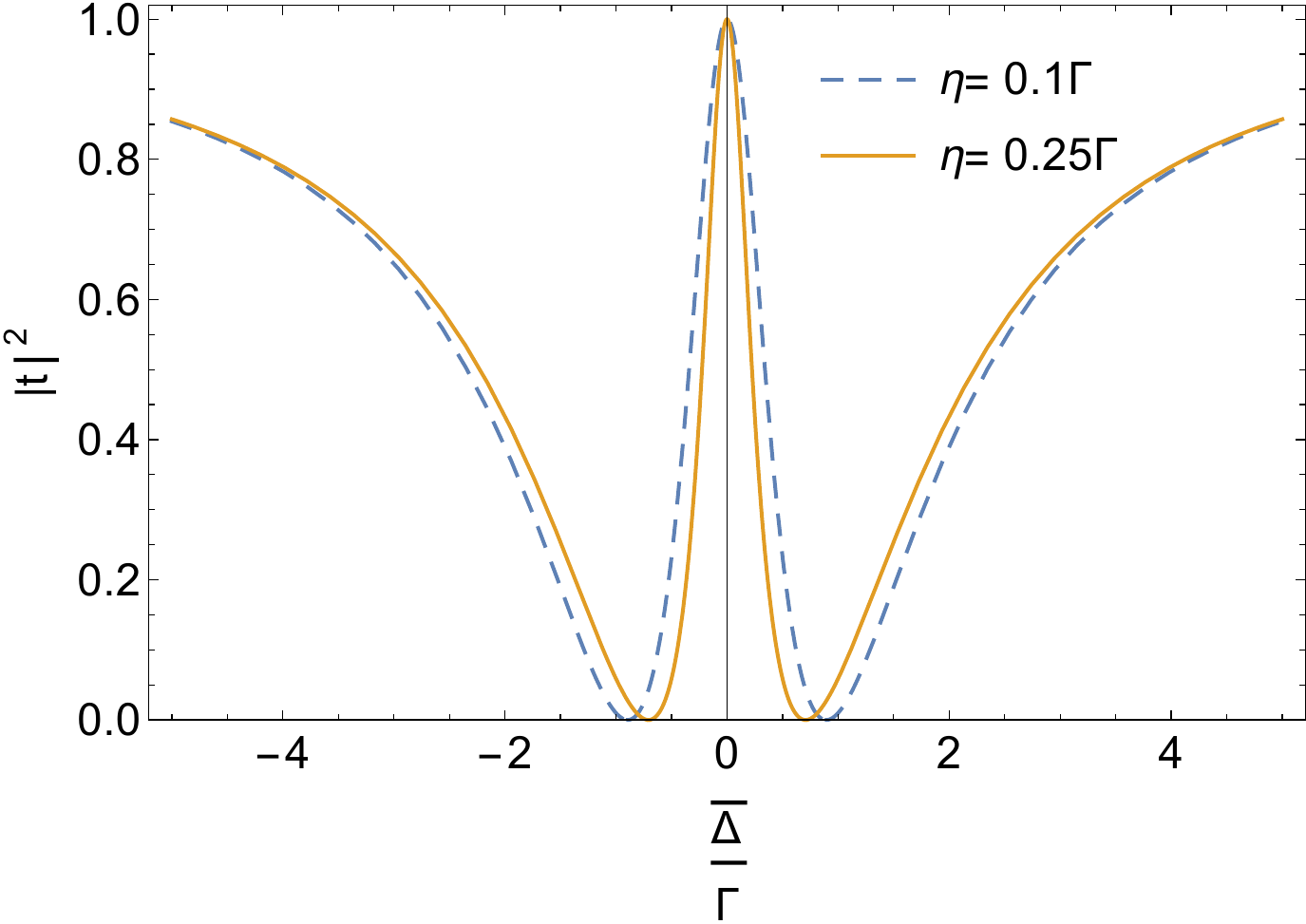} 
\caption{\small  {Transmission at $kL=n\pi$ for $\frac{s}{2}=\Gamma-\eta$ with $\eta=0.1\Gamma$ and $\eta=0.25\Gamma$.}} \label{f4} 
\end{figure}

The shrinking of the transmission width as $\frac{s}{2}$ goes below $\Gamma$ is clearly reflected in the transmission plots shown in Fig. \ref{f4}. Letting $\frac{s}{2}=\Gamma-\eta$, where $\frac{\eta}{\Gamma}\sim 10^{-1}$, we find that the pole $\overline{\Delta}_{+}^{(\text{p})}$ shrinks in width, with the relevant width given by $\bigg(1-\sqrt{\frac{2\eta}{\Gamma}}\bigg)\Gamma$. Thus, the transmission window becomes narrower as $\eta$ becomes larger.

\section{Transparency in a multiatom chain} \label{s4}

We now bring out some interesting features of the spectral behavior collectively induced by a chain of multiple emitters with non-identical detunings, corresponding to a spatial periodicity so chosen that $kL=n\pi$. In Ref. \cite{c37}, the analytical expressions for the spectral amplitudes were derived for a system of identical emitters for which the transfer matrices were also identical. Through a diagonalization procedure, the matrix product was calculated. However, $kL=n\pi$ is an exceptional point of the transfer matrices, since the eigenvalues coincide and become $(-1)^n$. Hence, diagonalization fails. However, in the special case when $kL$ is an integral multiple of $\pi$, one can derive exact analytical expressions for the relevant matrix product in \ref{E2}:
\begin{align}
\prod_{j=1}^N\mathcal{L}_j=(-1)^{nN}\begin{bmatrix}
1-i\Gamma\sum_{j=1}^{N}\delta^{-1}_{k (j)} & -i\Gamma\sum_{j=1}^{N}\delta^{-1}_{k (j)}\\
i\Gamma\sum_{j=1}^{N}\delta^{-1}_{k (j)} & 1+i\Gamma\sum_{j=1}^{N}\delta^{-1}_{k (j)}
\end{bmatrix}.
\end{align}

It is easy to check this result for $N=2$, and the general result for arbitrary $N$ can be be verified using the procedure of mathematical induction. We then have the following transmission and reflection coefficients:

\begin{align}
t=\frac{1}{1+i\Gamma\sum_{j=1}^{N}(\omega_k-\omega_j+i\Gamma_0)^{-1}}\hspace{1mm}, \notag\\
r=-\frac{i\Gamma\sum_{j=1}^{N}(\omega_k-\omega_j+i\Gamma_0)^{-1}}{1+i\Gamma\sum_{j=1}^{N}(\omega_k-\omega_j+i\Gamma_0)^{-1}}\hspace{1mm}.\label{E5}
\end{align}

\begin{figure}[!t]
 \captionsetup{justification=raggedright,singlelinecheck=false}
\centering
\includegraphics[scale=0.70]{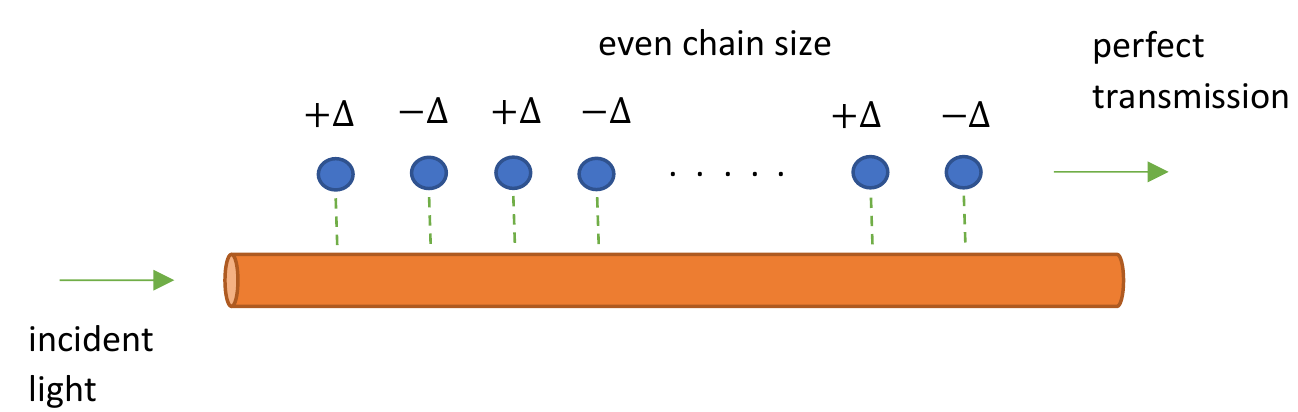} 
\caption{\small  {Even number of emitters with equal and opposite detunings assigned in pairs generates transparency. The order of the atoms is not important, so the arrangement shown here is just one of the possible permutations.}} \label{f5} 
\end{figure}

The collective effects due to emission from multiple periodically spaced emitters is clearly embodied in the aforementioned expressions.  The spectral dependence on the detunings has a close resemblance with that in the single-emitter scenario. The key factor that modifies the spectrum is $\sum_{j=1}^{N}(\omega_k-\Omega_j)^{-1}$, an additive effect of the inverse detunings pertaining to the individual emitters. The expression is, of course, not as simple for other choices of phase. As a further simplification, let us focus on the case where $\Gamma_0$ is small enough to be dropped from consideration. This, in principle, entails in the possibility of generating perfect transmission through suitable arrangements of the individual detunings. The condition for transparency ($r=0$ and $t=1$)is given by

\begin{align}
\sum_{j=1}^{N}\frac{1}{\Delta_{k(j)}}=0.
\end{align}

where $\Delta_{k(j)}=\omega_k-\omega_j$ is the laser detuning relative to the transition frequency of the $j^{\text{th}}$ atom. For a single emitter, this relation is clearly impossible to satisfy and therefore, a single atom in a waveguide does not lead to transparent behavior. One must have multiple atoms to enable the achievement of transparency. In a double-emitter scenario, where $N=2$, the condition translates to $2\omega_k=\omega_1+\omega_2$, which implies an exactly antisymmetric assignment of detunings to the two emitters. This is in line with what was highlighted in the previous section dedicated to the study of a two-atom chain (see Fig. \ref{f2}). 

For $N=3$, the corresponding constraint appears as a quadratic equation

\begin{align}
3\omega_k^2-2(\omega_1+\omega_2+\omega_3)\omega_k+\omega_1\omega_2+\omega_2\omega_3+\omega_3\omega_1=0,
\end{align}
with roots given by 
$$\frac{1}{3}\bigg[\omega_1+\omega_2+\omega_3 \pm\sqrt{\omega_1^2+\omega_2^2+\omega_3^2-\omega_1\omega_2-\omega_2\omega_3-\omega_3\omega_1}\bigg].$$ The discriminant can be re-expressed as $\frac{1}{2}[(\omega_1-\omega_2)^2+(\omega_2-\omega_3)^2+(\omega_3-\omega_1)^2]$, which, being a sum of squares, is strictly non-negative, and hence, the roots are real.

 In general, it is easy to see that for even number of emitters in the chain, it is always possible to make the system transparent if the atomic transition frequencies can be so adjusted that for any atom detuned by a certain amount, there exists another atom in the chain detuned by the same amount but in the opposite sense (Fig. \ref{f5}). In other words, for a chain of $2l$ atoms, an assignment of the frequency detunings $+\Delta_1$, $-\Delta_1$, $+\Delta_2$, $-\Delta_2$,... $+\Delta_l$, $-\Delta_l$, in no particular order, would give rise to transparency in the system. Such an asymmetric pairwise assignment of detunings lead to Fano minima in the reflection spectrum, which signifies a destructive interference between the reflected waves emanating from the emitters. Concomitantly, the transmitted waves constructively interfere, leading to perfect transmission. This extreme resonant inhibition of the reflection amplitude relative to the single-atom emission is a new phenomenon that is not observed at $kL=n\pi$ for a system of identically detuned emitters. 

\begin{figure}
 \captionsetup{justification=raggedright,singlelinecheck=false}
\centering
\includegraphics[scale=0.80]{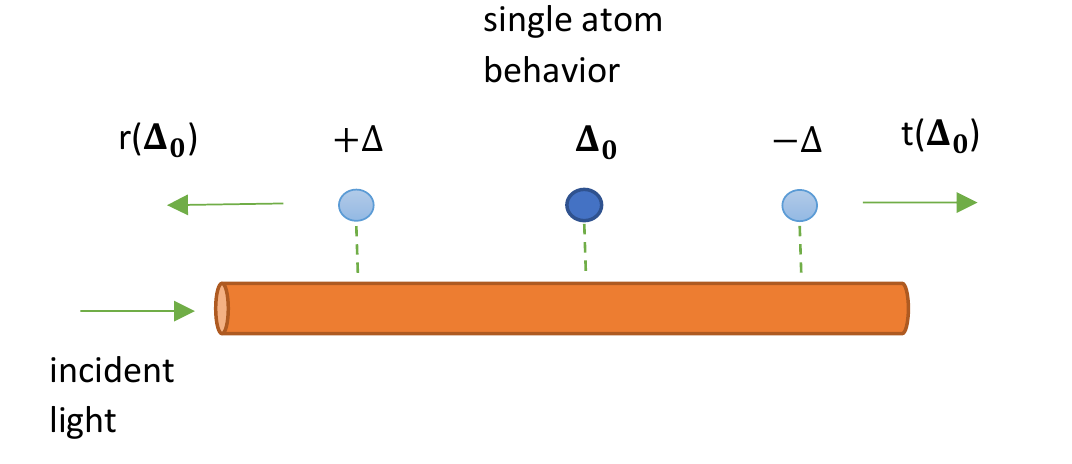} 
\caption{\small  {A system of three atoms, out of which two carry equal and opposite detunings $+\Delta$ and $-\Delta$. The odd one out (the middle one, in this figure) with a detuning of $\bm{\Delta_0}$ determines the spectral behavior, and no collective effects exist. This behavior transcends to the case of any odd number of emitters in the chain with a commensurate assignment of frequency detunings. When $\bm{\Delta_0}$$=0$ , the system behaves as a perfectly reflecting mirror.}} \label{f6} 
\end{figure}

On the other hand, if one has odd number of emitters in the chain, one can recover the single-atom emission spectra by assigning pairwise asymmetric detunings to any randomly chosen $(N-1)/2$ emitter pairs, leaving out a single atom. It then follows from Eqs. \ref{E5}, that the remaining atom completely determines the spectral characteristics. That is, if this particular atom has a transition frequency $\omega_0$, the transmitted spectrum due to the entire atomic chain reduces simply to
\begin{align}
t=\frac{1}{1+i\Gamma(\omega_k-\omega_0)^{-1}}\hspace{1mm},
\end{align}

which is identical to the transmission coefficient with just that single atom coupled to the waveguide (Fig. \ref{f6}). Stated differently, when an even number of atoms with a pairwise asymmetric assignment of frequency detunings are added, in a periodic fashion, to a single atom coupled to a waveguide, no discernible collective effects emerge. The order of this arrangement and therefore, the location of the odd atom are not important. This makes sense from the perspective of Fano interference, since the reflected waves from the appended atoms destructively interfere, while that from the residual atom effectively goes through unperturbed. As a consequence, if the odd atom is in resonance with the laser frequency, the system resembles a perfectly reflecting mirror, regardless of the frequency detunings of the other atoms.

Finally, in the event that all the atoms have identical frequencies, one can observe Dicke-type superradiant behavior due to enhancement of the reflection amplitude. If the atomic frequencies are all set equal to $\omega_0$, the corresponding transmitted spectrum is obtained to be

\begin{align}
t=\frac{1}{1+iN\Gamma(\omega_k-\omega_0)^{-1}}\hspace{1mm},
\end{align}

which pertains to a Lorentzian with a half-width of $N\Gamma$. This superradiant behavior observed for a collection of identical emitters, at $kL=n\pi$, was also derived in \cite{c37}. However, the possibility of controlling collective effects by tuning the individual atomic frequencies was not explored in that work. Having this added flexibility of adjusting the atomic frequencies brings out different types of interesting radiant behavior that one can observe, in principle. We do not merely encounter the possibility of superradiant reflection, but also come across new points of transparency. In particular, if we have an even number of emitters coupled to the waveguide, a pairwise asymetric allotment of detunings generates new Fano minima. For an odd chain size, we see how a similar assignment of frequencies to any $N-1$ of the atoms can lead to a complete disappearance of collective behavior and give rise to a spectrum governed entirely by the transition property of the leftover atom.

\section{Reciprocal behavior for $kL=n\pi$} \label{s5}

As has been discussed previously, in the context of a two-atom system, reciprocity entails for the choice of phase $kL=n\pi$. It follows, quite generally, from the expressions in \ref{E5}, that $kL=n\pi$ ensures perfect optical reciprocity for any arbitrary chain size. In fact, this choice of phase is both necessary and sufficient for reciprocity in both the reflection and the transmission. The fundamental property that brings about this reciprocal character is the commutativity between any two transfer matrices, i.e.
\begin{align}
[\mathcal{L}_j,\mathcal{L}_k]=0\hspace{2mm} \forall \hspace{1mm}j,k \label{E3}.
\end{align}

The commutation relation follows from  the form in \ref{E1}. As an essential implication of this, one finds that the matrix product $\prod_{j=1}^N\mathcal{L}_j$ is insensitive to the order in which the individual matrices are multiplied. Consequently, no matter what the order of the atoms is, one has the same transmission and reflection coefficients. Of course, this result is valid under the assumption that each emitter couples identically to the left as well as the right-propagating fields, as far as the two coupling strengths are concerned. It is easy to verify that Eq. \ref{E3} holds true for any arbitrary assisgnment of detunings if and only if $kL=n\pi$.

\section{Effect of dissipation into nonwaveguide modes on transparency} \label{s6}

\begin{figure}[!t]
 \captionsetup{justification=raggedright,singlelinecheck=false}
\centering
\includegraphics[scale=0.70]{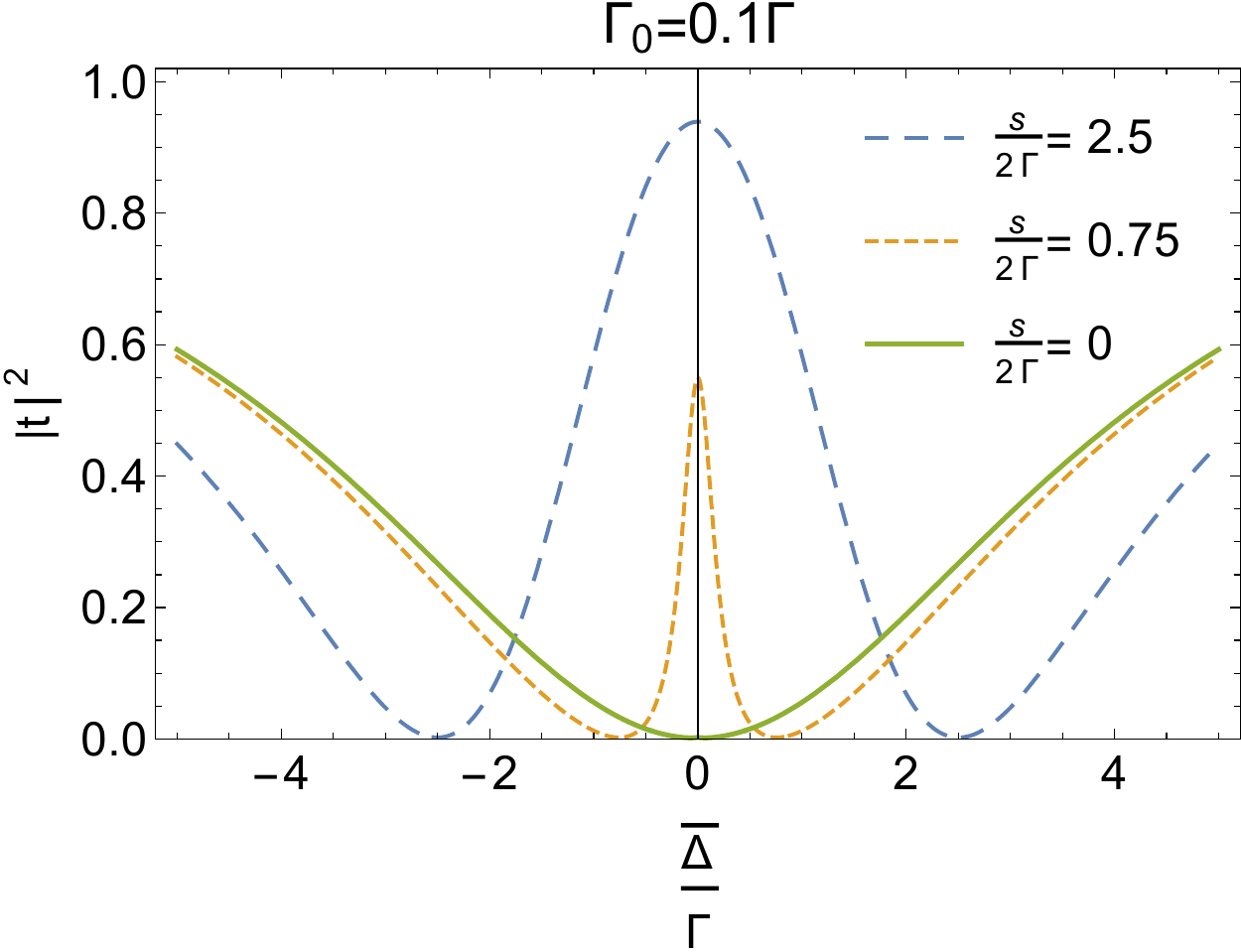} 
\caption{\small  {Effect of dissipation on the transmission of a two-atom system. If the dissipative channel is weak compared to the waveguide channel, the profile closely resembles the dissipation-free spectrum, except when the frequency mismatch between the atoms is smaller than or comparable to the rate of dissipation.}} \label{f7} 
\end{figure}

The simplistic results laid out in the preceding discussions in Secs. \ref{s3} and \ref{s4} hold only when $\Gamma_0$ is small enough to be ignored. However, we can look at more realistic scenarios with dissipation included ($\Gamma_0\neq 0$) and examine the effect of the same on those observations. For a two-atom chain, we discover that the behavior changes drastically depending on how the relative detuning between the atomic frequencies compares to this decay rate. Fig. \ref{f7} shows the plots for $\abs{t}^2$ vs. $\abs{\frac{\overline{\Delta}}{\Gamma}}$ for a good-quality waveguide with a weak dissipative channel ($\Gamma_0=0.1\Gamma$), for varying values of $\abs{\omega_1-\omega_2}$. It is observed that for sufficiently small values of the latter, the transmission peak almost disappears, whereas for large values, the height of the peak approaches unity. In other words, by adjusting the relative frequency detuning between the emitters, one can achieve either high opacity or high transparency around $\omega_k=\frac{1}{2}(\omega_1+\omega_2)$. For perfectly matched up atomic frequencies, i.e. $\omega_1=\omega_2$, one observes a diametrically opposite behavior as the two roots coincide - the central peak is replaced by a trough. This is a Dicke-type superradiant effect - for negligible decay, the transmission profile is a vertically inverted Lorentzian with a half-width of $2\Gamma$.

One can analytically understand this behavior by considering two specific regimes, (i) $s\ll 2\Gamma_0$ and (ii) $s\gg 2\Gamma_0$. At $\overline{\Delta}=0$, one obtains

\begin{align}
t=\frac{(\frac{s}{2})^2+\Gamma_0^2}{(\frac{s}{2})^2+\Gamma_0(\Gamma_0+2\Gamma)}\hspace{1mm}.
\end{align}

For small relative detuning between the atoms, i.e. $s\ll 2\Gamma_0$, the approximate form is given as $\abs{t}^2\approx \frac{\Gamma_0^2}{4\Gamma^2}$, which is vanishingly dimunitive, as long as the decay rate is much less than $\Gamma$. In the opposite scenario when $s\gg 2\Gamma_0$, we get $\abs{t}^2\approx 1-\mathcal{O}(\frac{8\Gamma\Gamma_0}{\delta^2})$. Thus, a fairly high degree of transparency can be achieved by specifically working with a large relative detuning $\abs{\omega_1-\omega_2}$.

For a system of even number of emitters, in which half of them have detuning $+\Delta$ whereas the other half have detuning $-\Delta$, the transmission goes as

\begin{align}
t=\frac{\Delta^2+\Gamma_0^2}{\Delta^2+\Gamma_0(\Gamma_0+N\Gamma)}\hspace{1mm}.
\end{align}

For $\Delta\ll\Gamma_0$, $\abs{t}^2\approx (\frac{\Gamma_0}{N\Gamma})^2$, which testifies to highly reflecting behavior. However, when $\Delta\gg \Gamma_0$, we have $\abs{t}^2\approx 1-\mathcal{O}(\frac{4N\Gamma\Gamma_0}{\Delta^2})$, implying near-transparency. The situation here is reminiscent of EIT where perfect transparency emerges in the absence of dissipative transitions \cite{harris1991,eitreview}.

Similarly, when there are odd number of emitters, with $(N-1)/2$ emitters each having a detuning of $+\Delta$ and $(N-1)/2$ other emitters each detuned by $-\Delta$, one has, for the transmission coefficient

\begin{align}
t=\frac{1}{1+\frac{(N-1)\Gamma\Gamma_0}{\Delta^2+\Gamma_0^2}+\frac{i\Gamma}{\omega_k-\omega_0+i\Gamma_0}}\hspace{1mm},
\end{align}

where $\omega_0$ is the frequency of the remaining atom. When $\Delta$ is large compared to $\Gamma_0$, one can discern a re-emergence of single-atom behavior.

\section{Conclusions} \label{s7}

To summarize, we have thrown light on new possibilities that emerge in relation to the collective effects of a chain of atoms side-coupled to a waveguide when the interemitter separation is fixed to satisfy $kL=n\pi$, where $n$ is an integer. For a chain of $N$ atoms, we have demonstrated the emergence of new Fano minima (transparency points) in the reflection spectrum for negligible dissipation. When $N$ is even, we have seen how transparency can be generated by assigning equal and opposite detunings to the atoms in pairs, while for odd $N$, we have highlighted the possibility of reproducing single-atom behavior through a similar assignment, so that the odd one out completely determines the emission spectrum. A system of identically detuned emitters, on the other hand, demonstrates superradiant behavior. We have also shown that the optical system demonstrates reciprocal behavior with respect to both transmission and reflection. In general, the system turns out to be insensitive to the order in which the atoms are arranged. Finally, it has been demonstrated, both analytically and graphically, that when dissipation into nonwaveguide modes cannot be neglected, one can still produce highly transparent behavior by implementing a considerable disparity in the atomic transition frequencies. For a small mismatch in the frequencies, one, however, observes predominantly opaque behavior in its place.\\


\begin{acknowledgments}

D. M. is supported by the Herman F. Heep and Minnie Belle Heep Texas A\&M University Endowed Fund held/administered by the Texas A\&M Foundation.

\end{acknowledgments}

\end{document}